\documentclass[journal]{IEEEtran}

\usepackage{amsmath}
\usepackage{amsthm}
\usepackage{amssymb}
\usepackage{float}
\usepackage{mwe}
\usepackage[ruled,vlined]{algorithm2e}

\begin{document}

\title{Attention-effective multiple instance learning on weakly stem cell colony segmentation}

%\author[1\authfn{1}]{Muthu Subash Kavitha}
%\author[2]{Novanto Yudistra }
%%\author[3]{Jeny Rajan }
%\author[4]{Takio Kurita }

\author{Novanto Yudistira, Muthu Subash Kavitha, Jeny Rajan, Takio Kurita

        %Novanto Yudistra,
        %Jeny Rajan,
        %Takio Kurita}% <-this % stops a space

\thanks{N. Yudistira is with Intelligent System Lab, Faculty of Computer Science, Brawijaya University, Indonesia
e-mail: yudistira@ub.ac.id}

\thanks{M.S. Kavitha is with School of Information and Data Sciences, Nagasaki University, Nagasaki, Japan
e-mail: kavitha@nagasaki-u.ac.jp}
 
\thanks{J. Rajan is with Department of Computer Science and Engineering, National Institute of Technology Karnataka, Surathkal, India}
 
\thanks{T. Kurita is with Graduate School of Advanced Science and Engineering, Hiroshima University, Hiroshima, Japan
e-mail: tkurita@hiroshima-u.ac.jp}
}
% Include full affiliation details for all authors
%\affil[1]{School of Information and Data Sciences, Nagasaki University, Nagasaki, Japan}
%\affil[2]{Faculty of Computer Science, University of Brawijaya, Indonesia}
%\affil[3]{Department of Computer Science and Engineering, National Institute of Technology Karnataka, Surathkal, India}
%\affil[4]{Graduate School of Advanced Science and Engineering, Hiroshima University, Hiroshima, Japan}

%\corraddress{Muthu Subash Kavitha, School of Information and Data Sciences, Nagasaki University, Nagasaki, Japan}
%\corremail{kavitha@nagasaki-u.ac.jp}

%\runningauthor{Kavitha et al.}

%\begin{frontmatter}
\maketitle

\begin{abstract}

The detection of induced pluripotent stem cell (iPSC) colonies often needs the precise extraction of the colony features. However, existing computerized systems relied on segmentation of contours by preprocessing for classifying the colony conditions were task-extensive. To maximize the efficiency in categorizing colony conditions, we propose a multiple instance learning (MIL) in weakly supervised settings. It is designed in a single model to produce weak segmentation and classification of colonies without using finely labeled samples. As a single model, we employ a U-net-like convolution neural network (CNN) to train on binary image-level labels for MIL colonies classification. Furthermore, to specify the object of interest we used a simple post-processing method. The proposed approach is compared over conventional methods using five-fold cross-validation and receiver operating characteristic (ROC) curve. The maximum accuracy of the MIL-net is 95\%, which is 15\% higher than the conventional methods. Furthermore, the ability to interpret the location of the iPSC colonies based on the image level label without using a pixel-wise ground truth image is more appealing and cost-effective in colony condition recognition.

% Please include a maximum of seven keywords
\end{abstract}
%\end{frontmatter}

\section{Introduction}

\IEEEPARstart{I}{nduced} pluripotent stem cells (iPSC) can self-renew infinitely and generate into  every human body's cell type. The iPSCs are helpful to substitute deteriorated tissue of the human body and thus it is highly demanded clinical drug development \cite{takahashi}. To realize reliable and secured tissue regeneration, it is essential to determine the conditions of the cells during their culture. Identifying the good quality cells and colonies (cluster of identical cells) for subsequent treatment therapy is generally observed by the eye in terms of the morphological features, such as colonies with a densely packed cell appearance and almost a well-defined edge. On the contrary to the morphology of the excellent quality cells, harmful quality colonies are detected.  However, manual evaluations of cell conditions highly rely on human experts and cost errors  \cite{fan}. Furthermore, the assessment of a massive amount of cell conditions in culturing is tedious and laborious. Hence non-invasive automatic classification technique would benefit from tracing large numbers automatically interestedly without any classification errors.

Induced pluripotent stem cells (iPSC) can self-renew infinitely and generate into  every human body's cell type. The iPSCs are helpful to substitute deteriorated tissue of the human body and thus it is highly demanded clinical drug development \cite{takahashi}. To realize reliable and secured tissue regeneration, it is essential to determine the conditions of the cells during their culture. Identifying the good quality cells and colonies (cluster of identical cells) for subsequent treatment therapy is generally observed by the eye in terms of the morphological features, such as colonies with a densely packed cell appearance and almost a well-defined edge. On the contrary to the morphology of the excellent quality cells, harmful quality colonies are detected.  However, manual evaluations of cell conditions highly rely on human experts and cost errors  \cite{fan}. Furthermore, the assessment of a massive amount of cell conditions in culturing is tedious and laborious. Hence non-invasive automatic classification technique would benefit from tracing large numbers automatically interestedly without any classification errors.

Several automated techniques have been developed to classify various conditions of iPSCs \cite{Yuan, kavitha, jout}. Several research works using digital image processing techniques exploiting preliminary filtering and thresholding to detect the shape of the objects of the colonies \cite{kavitha, Zhang2009}. However, the feature assessment using image analysis techniques depends on prior parameters and manual interactions, prone to large-scale assessment errors. Furthermore, the morphology of colonies is dynamically changed in subsequent reprogramming stages. Thus prior parameter setting approaches were not appropriate for evaluating the colonies \cite{NAGASAKA201741, Kato}. In order to alleviate manual interaction, few approaches used machine learning techniques. However, machine learning methods relied on hand-crafted microscopic morphology-based and texture-based features of colonies to classify cell conditions \cite{jout, Stumpf, CompBio}. Specifically, hand-crafted features-based support vector machine (SVM) models were commonly applied and produced satisfactory results for the classification of conditions of colonies \cite{Ray, subash}.

Recently deep learning methods are extensively used in detecting cell images because of the ability to recognize the changes and development of stem cells without manual interventions \cite{WAISMAN, Moe}.  The open-source package and Xception network were effective in differentiating the types of neural stem cells \cite{NatCom}. A vector-based convolutional neural network (V-CNN) was developed and matrix transformation is added as a pre-processing layer in the two-dimensional CNN \cite{kavitha}. The authors in \cite{kavitha} claimed that the V-CNN produced better performance than SVM for the classification of colonies. A simple LeNet architecture with an image processing algorithm efficiently derived cell types from iPSCs with high performance \cite{KUSUMOTO}. However, the methods mentioned above highly relied on the number of pre-processing ways to locate most related features for iPSC colony classification. The pre-processing steps are often problem-specific and required prior parameter settings, which is not always appropriate for evaluating the variations in iPSCs heterogeneity. Thus we intend to develop a single model without pre-processes for reducing the risk of biased results and inconsistencies for colony conditions evaluation.

We used a customized version of the popular encoder-decoder based U-net architecture \cite{ronneberger2015u}. Previously, several biological imaging tasks have been utilized U-net or attention mechanism for segmentation due to its ability to capture coarse-to-fine structures \cite{yudistira2020prediction, du2020medical,oktay2018attention}.
Differently in this study, we proposed to use an end-to-end MIL-net without pre-processing that implements local connectivity patterns between the neurons of the adjacent layers and average pooling for the attention features at the end of the architecture. MIL is a specific type of supervised learning, where instances are grouped into sets, termed as bags, and labels are only given at the bag level and not for each individual instance level. In our case, MIL attempts to discover the target variable from the instances of sparse and dense patterns of stem cells. The instances of stem cells are extracted through several convolution layers and transformed into a low dimensional space. There it can generate a single bag level representation using average weighted pooling with highest attention to show landmarks of stem cell region as well as classifies the bag into good or bad colony image. The classical global pooling methods can only detect approximate pixel location, and thus, global weighted average pooling was used to evaluate the pixel level localization \cite{GlobalWA}. A fully convolutional neural network trained with fewer ground truth bounding boxes and many image-level labels was found to be effective in locating the pixel-level objects on the benchmark datasets \cite{GlobalWA}. Attention gating as Sononet was used in VGG or U-net to detect salient regions on the medical images \cite{AttentionGN}. Attention mechanism using GRADCam in U-net with logistic regression classifier enhanced Alzheimer’s disease classification \cite{kavitha2019multi}. A modified 25-layers of U-net was effectively used to diagnose cardiac arrhythmia based on the electrocardiographic signals \cite{AutomatedBA}. A weakly-supervised approach using feedback CNN and global average pooling with binary labels was used to locate the satellite images \cite{FeedbackNN}. The attention mask generated from the attention U-net improved the iris region detection \cite{LIAN}.

Motivated by the studies mentioned above, we intended to utilize the effects of the MIL through a weakly supervised approach, where the binary image-level label (colony with dense cells as good /sparse cells as bad) is given to the group of instances. However, the aforementioned MIL-based CNN architectures extracted local to global features from the multiple non-linear layers limiting the performance by insisting on the intensity profile with shape features. In order to improve that, we intend to push the local colony structure information in the MIL-net by highlighting the essential features for colony conditions. MIL is used to determine a sample when all of the instances from a sample must be taken into account without any specific pixel label to the each of the instances. In our case, MIL attempts to discover the target variable from the instances of sparse and dense patterns of stem cells by extracting the feature maps through several convolution layers and transformed into a low dimensional space. There it can generate a single bag level representation using average weighted pooling and classifies the bag into good or bad colony image.
Furthermore, supervised learning demanding a large amount of annotated images, which are tedious and time-consuming. Alternatively, the proposed approach based on a U-net-like structure to predict the pixel-level segmentation with the boundaries of the colonies without a finely-labeled sample is promising in cell detection. 

The  contribution  of  this  study  can  be  summarized  as  follows:

1)  Proposes a multiple instance approach in form of weakly supervised for iPSC colony segmentation and colony conditions classification based on Unet-like architecture in an end-to-end manner without using finely-labeled samples.

2) Involves simple post-processing in the learning output to automatically specify region of interests and removes the unwanted pixel localizations as false positives.

3) Compares the performance of the proposed framework over architectures that includes U-net with fully connected layer (hereafter termed as baseline), patch-based shallow U-net, ResNet-50, deep V-CNN and SVM. 

4) Investigates the performance of the proposed model using five-fold cross-validation. 

5) Evaluates the performance of all architectures by using mean accuracy, precision, recall, F-score and receiver operating characteristic curve (ROC) measures.

\section{Materials and Methods}
\subsection{Dataset}
This study included a set of 94 images of iPSC colonies. Out of 94 datasets, 60 were maintained as described elsewhere \cite{okita} and 34 were received from American Type Culture Collection. The details of gathering the iPSC colonies and phase contrast microscopic image collection settings are explained in our previous study \cite{kavitha}.

\subsection{ Proposed MIL for discriminative cell patterns for colony condition }
Colony condition recognition is a typical binary image classification problem for a learning algorithm. Consider X is the input image, and N represents the total
number of classes.

\begin{figure*}
\centering
\includegraphics[width=1.0\textwidth]{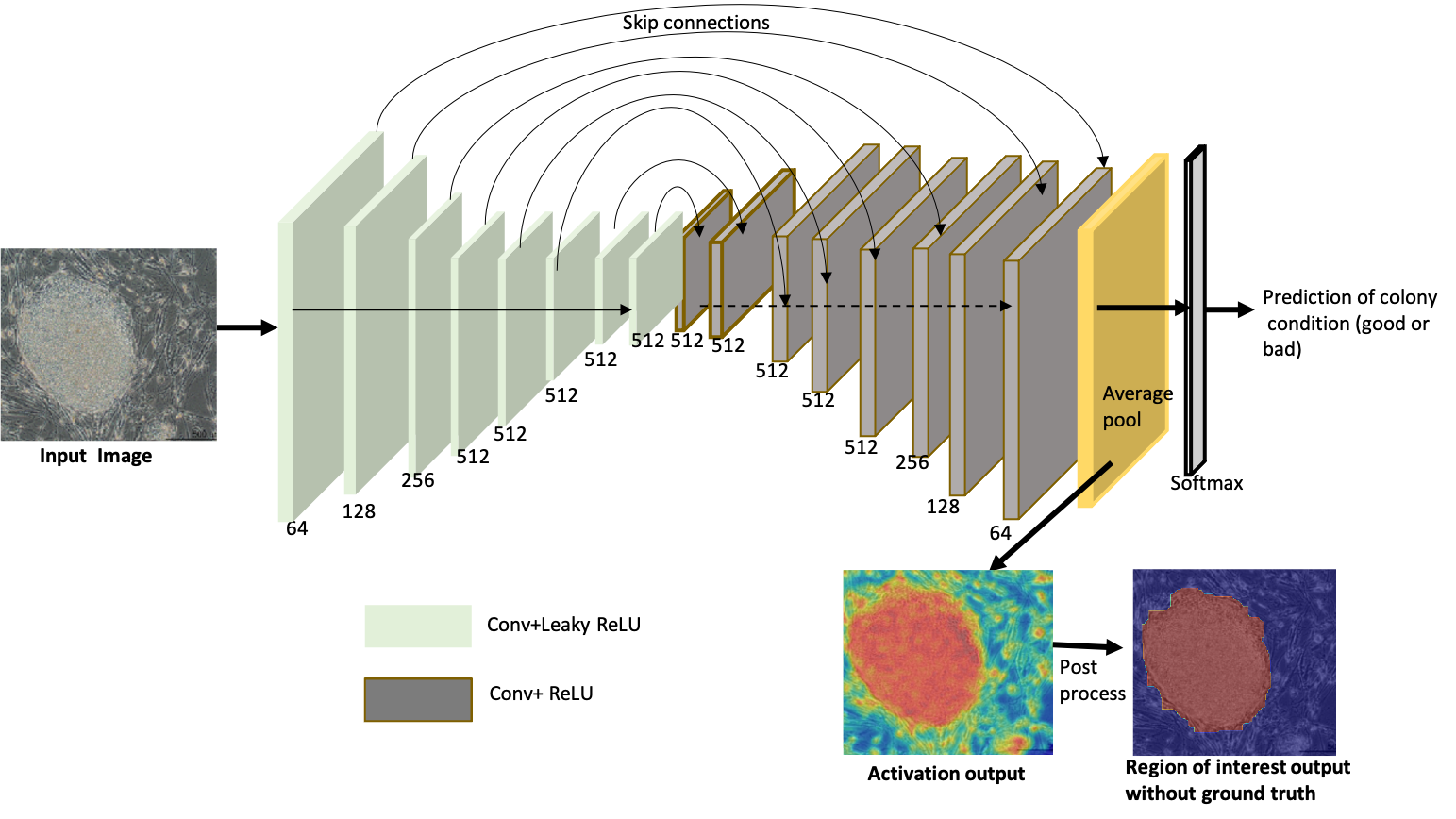}
\centering
\caption{\label{fig:Picture 1ipsc} Proposed inference-effective learning network for induced pluripotent stem cell colony conditions classification.}
\end{figure*}

Hence, L $\in $ $\{1, ..., N\}$ is the corresponding class label of X. The training algorithm proposes to find a function $f$ : X $\rightarrow $ L. In conventional image classification frameworks for colony condition detection, $f$ is often defined as G(E(X)), where E(X) and G(.) indicate the feature extractors and classifiers, respectively. Furthermore, in the concepts of MIL-based CNN architectures, $f$ consisted of multiple non-linear layers with convolutional layers, each followed by pooling and one fully connected and softmax classifier to extract local to global features. However, in the process of colony condition recognition, the local definite structure information in the cell colony is often asymmetrically distributed. Hence, learning global features from the conventional multiple non-linear layers based-CNN limited the performance in distinguishing the colony condition. Furthermore, it demands the intensity profile along with the shape features. To resolve this issue, we constructed an end-to-end U-net like deeper CNN architecture adopted to MIL framework by pushing the local colony structure information for colony condition detection. More of interest,  automatically partition the informative local colony contour according to the conditions of the colonies without finely labeled samples are promising in cell detection. 

The proposed framework fits into MIL criterion and  consists of train encoder and decoder stages (Figure 1). We trained our proposed architecture independently from scratch. No pretraining or transfer learning was used in any of our experiments in this study. It fits into the weakly supervised learning in which input data is labeled as good/bad called bags \cite{Zhi, Foulds}. The bags can be termed as relation between instances. Hence the positive bag included positive instances and negative bag included negative instances with label $Y=\{y_1,y_2,...,y_n\}$ of +1 and -1, respectively. Then MIL label follows the equation as,

\begin{equation}
Y = 
  \begin{cases}
  +1, &\text{if } \exists y_i : y_i = +1;\\
  -1, & \text{if } otherwise.
  \end{cases}
\end{equation}

We used the criterion of bags and instances in the MIL settings. In this study, a single channel $250 \times 250$ pixels size of good and bad quality iPSC colony images are used with their categorical labels to train the model.  

Image level labels or bag labels of input data is used to generate pixel locations or instance labels of the cell area. The bag labels consist of two classes of images in this study.  Furthermore, in this study we did not label the exact location or each pixel in the stem cell regions, instead image-wise good and bad labels are used to train the network. However, each pixel locations were generated from the network weights that trained after the backpropagation by using average activation.  There, it can automatically find the region of interest by visualizing the stem cell colonies and it is indicated as a weak segmentation in this study. Furthermore, the highest attention that contributing the cell region enhance the performance in classifying the conditions of the good and bad colonies. The steps involved in the proposed MIL-net are; classification of the classes of good or bad iPSCs until convergence and weakly supervised segmentation of the cell colonies retrieve from the last convolutional layer of the network.

\subsection{Multiple instance  classification  }
In this study we propose a U-net-like CNN architecture that automatically finds regions of interest and differentiates between different cell conditions of iPSC, such as good and bad, in an end-to-end fashion (Figure 1). The feature maps of decoder is concatenated with the feature maps that are skipped from the encoder through the skip connections and thus it can able to retrieve the full spatial resolution at the network output. Furthermore, it added an average pooling layer at the end of the network layer to enhance the classification capacity of the CNN. The proposed architecture is designed by large receptive fields of the output neurons, essential for multiple instance classifications. The new addition of average pooling and fully connected layer at the final layer complement their plain counterparts in the classical U-net architecture. Therefore, it is trained by leveraging and back-propagating the network to give classification results. 

The gray scale images of iPSC input images are fed into the network. 
We also used augmentation images involved with vertical or horizontal flip and rotation of 90, 180 and 270$^{\circ}$ degrees.
The network  implemented $m \times m$ convolution filters for features extraction and dimension reduction. The softmax is used to find the probability of the colony conditions followed by the average pooling and fully connected layer. The encoder and decoder structure combines the feature maps. The encoder consists of eight layers of convolutions and leaky ReLU. Each convolutional layer has $4\times4$ kernels with stride $2$. The encoder part starts with 64-dimensional features or channels and it increases until to reach 512 feature maps. In the encoder, the input $x$ convolved with filter of $w^c$ and residual bias $b_{cl}$, where $c$ is channel number and $l$ is layer number, before fed into non-linear activation function of $f$ (leaky ReLU). It is defined as 

\begin{equation}
y_{ij}^l=f \bigg(\displaystyle \sum_{p=0}^{m} \displaystyle \sum_{q=0}^{m} w_{p,q}^cx_{(p+i)(q+j)}^{l-1}+b_{cl} \bigg)
\end{equation}

The convolutions are done spatially in 2-dimensional space, where $p$ and $q$ are width and height of the input, respectively. The previous output layer $x^{l-1}$ is convoluted and activated to produce the activation output of $y^l$.

The decoder path consists of eight up-sampling convolutions called deconvolution layers \cite{yang2021deconvolution} and ReLU activation functions. Drop out operation is added on first three up-sampling convolutions that is after  ReLU. It maximizes feature maps by 4, and minimizes the number of features dimension by half.
If there is any negative activations,  ReLu returns zero and hence the gradient become zero for all the inputs to the following layers. However, Leaky ReLu returns very small value for any negative inputs. Hence, we used Leaky ReLu in the encoder and ReLu in the decoder. The output of the last up-sampling layer is passed into the average pooling layer followed by fully connected layer.
The high-level feature vectors of the last convolution layer are derived from low to high layers are passed into the average pooling. Average pooling helped to reduce the number of parameters as well as make the features invariant to varying locations, rotations, and scales that beneficial for generalization. Hence the precise attention and compact features derived from the average pooling are flows into the softmax cross entropy for classification can be capable to maintain the most relevant features for stem cell region and that enhance the network efficiency in categorizing the classes of colony conditions. The attention of the average pooling is used to localize the region of interest for visual interpretation and termed as weakly localization of the stem cell regions in this study.

 The filters in the decoder are also trainable parameters. The output $x$ of the previous layer is transposed and convoluted with filter of $w^{c}$ with the bias value of $b_{j}$ before nonlinear activation function of $f_{1}$. The total number of trainable parameters obtained from eight convolution layers of encoder and seven deconvolution layers of decoder are 54,653,008. The last convolution in the decoder is used for the reconstruction to the original size.

%\begin{equation}
%z=f_{1}\left\{w_{j}^{T}y+b_j \right \}
%\end{equation}

 Furthermore, the direct  localization via classification using the MIL-net based CNN method is evaluated by removing the average pooling with three fully connected layers and patch-based shallow number of layers for the conditions of the iPSC colony. The input and the number of neurons in the three fully connected layers are 5000, 1000 and two, respectively, corresponding to the good and bad conditions of colony. For comparison, additionally we built a MIL-net using patch based input. It is evaluated to reduce the computational burden. It included shallower layers of three convolutions and deconvolutions with average pooling at the final layer before the softmax.

 \subsection{Weakly supervised visualization}
 We used the softmax cross-entropy loss function by learning from image-level labels to train the proposed network. The proposed MIL-net directly learns the input and output relationship of different colony conditions. The classification of the colonies is learned from the network weights that learn after the convergence. The intermediate activation output of the last convolutional layer is used to visualize the localization of the region of interest. The weakly supervised learning of the proposed network visualizes the texture features of the colonies by minimizing the irrelevant neuron activation. The output investigates distinctive textures that identify the condition of the colonies.

\subsection{Post-processing}

The colony region obtained from the learning is not clearly delineated and still included some outliers which are not important for the detection. Particularly the neural network is highly uncertain with less the number of training data and thus this condition often occurs. Hence a simple post-processing step is needed to remove the unwanted pixel localizations. In order to achieve this, we used the morphological operation such as an opening that keeps the largest localized object visible. The remaining objects are removed as false positives. The opening operation consists of erosion and then dilation with structuring element or kernel. We used rectangle-based kernel of $z$ with $20\times20$ in size by experiment. The formulation of opening operation is shown as follows:

\begin{equation}
S \circ z = (S \ominus z) \oplus z
\end{equation}

Where S and z are input and kernel, respectively. The parameter $z$ performs erosion morphological operation ($\ominus$) on input $S$. And then dilation morphological operation ($\oplus$) is performed on $S$. These consecutive operations removed the small noisy artifacts and keeps iPSC colony region as the region of interest.
 
\section{Experimental setup}
\subsection{Experimental evaluation settings}
The proposed MIL-net classification method is applied to iPSC of good and bad colonies to evaluate its utility and effectiveness. Out of 94 images,  the number of good and bad colonies used in this study are 54 and 40,  respectively. The dataset is randomly partitioned into 74 training and 20 testing images without using any same instances in both train and test set. The good and bad colonies are 44 and 30, respectively in the training and 10 and 10, respectively in the testing set. To avoid overfitting we applied several regularizer methods throughout the network during training phase. In our architecture, we used dropout at the first three deconvolutional layers, batch normalization at Conv1 to DeConv7, ReLU for all layers, and weight decay. Dropout is important to prevent co-dependent neuron units and thus only the key properties are selected within thinned networks. To prevent covariate shift which occurs when the distribution between training and testing data are different while the conditional label distributions are the same, batch normalization is utilized. Finally, Leaky ReLU and ReLU activations are applied across layers to guarantee sparseness by removing unnecessary negative values which is beneficial for generalization. The proposed approach is compared over baseline U-net, patch-based shallow Unet, ResNet50, deep V-CNN and SVM methods. In patch-based shallow U-net we used patches of input as ($48 \times 48$) to train the network. Different from U-net, ResNet-50 only considers using encoder as end to end learning of image with skip connections and blocks \cite{he2016deep, Jifara}. And it can overcome vanishing gradient problems of deep network and thus it can allow to train with deeper layer without severe over-fitting. The performance of all architectures used in this study is compared using accuracy, precision, recall and F1-score.Additionally, the performance of the proposed approach is evaluated using five-fold cross validation by randomly splitting train and test using five times without repeating the same instances in each fold.Furthermore, we used receiver operating characteristic (ROC) curve to evaluate the performance of the architectures in classifying the colony qualities. We used the same training and testing splits for all the methods compared in this study. All the methods used in the 5-fold cross validation experiment are used the same train and test set splits for all the folds. The best hyper parameters are selected heuristically using the proposed model. The networks are trained using Adam optimizer with starting alpha (learning rate of Adam), beta, and weight decay of 0.0001, 0.5, and 0.000001, respectively. To avoid local minima, the alpha value of Adam is decreased by multiplying it with 0.9 for every 20000 iterations out of 300 epochs. The learning rate is dynamically and gradually reduced from 0.0001 to 0.00001 which make the loss smoothly decreased overtime. Thus, the final learning rate of 0.00001 is achieved and that lead smooth convergence. By experiment we set the learning rate of 0.001 to the baseline U-net. All the architectures were implemented in Python using the Chainer framework.

\section{Results and discussions}

As shown in Table \ref{tab:1}, the proposed MIL-net learning architecture outperforms all other architectures experimented in this study. The accuracy of the MIL-net is higher than the baseline, patch-based shallow-net and ResNet-50 by 5.0 \%, 35.0 \%, and 15.0 \%, respectively. Patch based segmentation or classification becomes alternative to network to learn via data augmentation. It increases variation thus network can learn more. However, in this study patch-based shallow network can not perform well because patch based is failed to learn global texture.

\begin{table}[bt]
\caption{\label{tab:1}Performance comparison of the MIL-net with other architectures}
\scalebox{0.9}{
\begin{tabular}{ | c | c | c | c | c |}
\hline
\textbf{Architectures} &  \textbf{Accuracy} & \textbf{Precision} & \textbf{Recall} &
\textbf{F-Score}
\\
\hline
MIL-net  &\textbf{0.95}  & \textbf{0.99} & 0.89 &\textbf{0.95} \\
Baseline  & 0.90 & \textbf{1.0} & 0.8 & 0.89\\
Shallow-net & 0.60 & 0.57 & 0.80 & 0.70\\
ResNet-50  & 0.80 & 0.75 & \textbf{0.90} & 0.82\\
\hline
V-CNN & 0.93 & 0.90 & \textbf{0.90} & 0.90\\
SVM & 0.83 & 0.84 & 0.82 & 0.82\\

\hline
\end{tabular}}

\end{table}

Though, the experimental results of the ResNet-50 is high with reasonable accuracy, the spatial pooling nature of the encoder make the network difficult to visualize. While compared to the SVM and V-CNN, the MIL-net outperformed by 12.0 \% and 2.0 \%, respectively. and does not require pre-training and pre-processing techniques to extend the localization of cell regions through the classification.  Furthermore, the proposed approach has lower number of parameters than the baseline and yields high F-score with the value of 95.0 \% compared to 89.0 \%. of baseline.

As shown in Table \ref{tab:2}, using an average of five-fold cross validation, the MIL-net is still the best in terms of recall with 96.0 \%.  The baseline shows good performance in terms of F-score. However, it has higher number of parameters than the MIL-net. Thus the proposed structure is still considered to be beneficial.

\begin{figure}[bt]
\centering
\includegraphics[width=7cm]{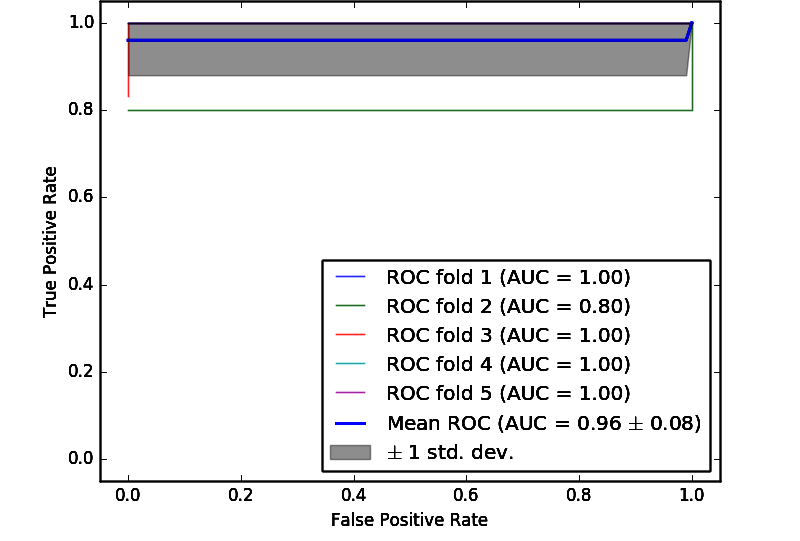}
\centering
\caption{\label{fig:roc_avg_pool}Evaluation of the receiver operating characteristic curve of the proposed approach based on five-fold cross validation.}
\end{figure}

Figure \ref{fig:roc_avg_pool} shows the ROC graph of the proposed network based on five-fold validation. It shows that the probabilities of different thresholds produce almost similar accuracy based on the network output. Furthermore, our data set used both separate as well as combined not more than two colonies in the dataset. The stem cell colony condition detection is not intended to separate the boundary of the combined colonies and hence, the performance of the MIL-net is not affected with the combined colonies. 
\begin{figure}[bt]
\centering
\includegraphics[width=6cm]{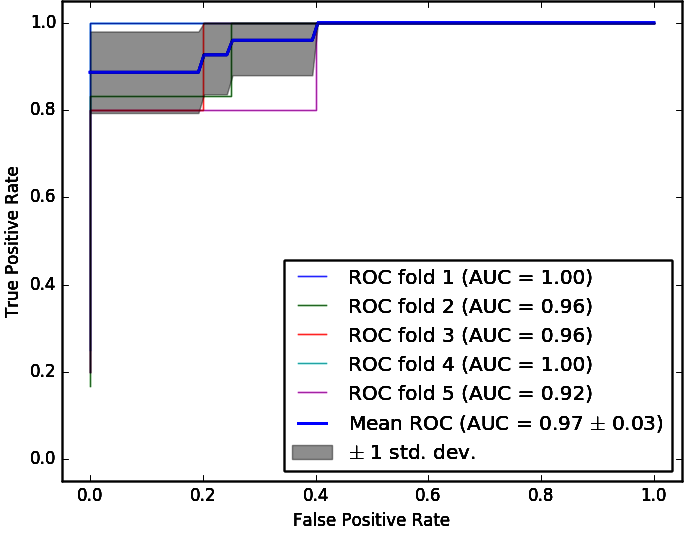}
\centering
\caption{\label{fig:roc_fc}Evaluation of the receiver operating characteristic curve of the baseline based on five-fold cross validation.}
\end{figure}
Compared to MIL-net as in Figure \ref{fig:roc_avg_pool}, the baseline as in Figure \ref{fig:roc_fc} is better in terms of mean ROC of five-folds with AUC of 97.0\%. However, in terms of the ROC graph, MIL-net outperforms as revealed from figure.  

\begin{figure}[bt]
\centering
\includegraphics[width=6cm]{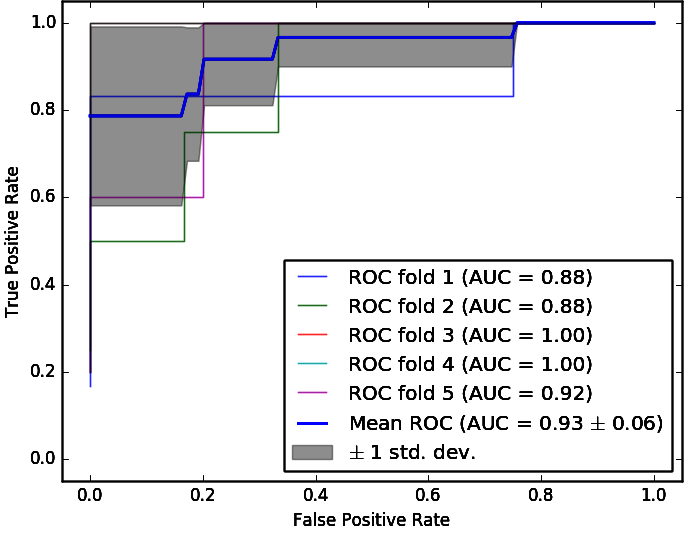}
\centering
\caption{\label{fig:top_down} Evaluation of the receiver operating characteristic curve of the ResNet-50 based on five-fold cross validation.}
\end{figure}

Figure \ref{fig:top_down} shows the ROC of the ResNet-50 in classifying the colony conditions using five-fold cross validation. It describes lower performance than the proposed and baseline with the mean AUC of 93.0\%. 
The proposed MIL-net classify the colony quality followed by automatic localization of cell areas which is different from the traditional  cell classification that performed classification after feeding the localized cell regions from several pre-processing steps. The approach used large series of mouse embryonic stem cells based on various architectures of deep CNN with annotations revealed 99.0\% accuracy in differentiating two different types of cells [13]. The detection rate of neuron in neural stem cells using Xception network was 92.0\% [15]. The stem cell differentiation using simple and

\begin{table}[h]
\centering
\caption{\label{tab:2}Performance comparison of the MIL-net using five-fold cross validation}
\scalebox{0.9}{
\begin{tabular}{| c | c | c | c | c |}
\hline
\textbf{Architectures} &  \textbf{Accuracy} & \textbf{Precision} & \textbf{Recall} &
\textbf{F-score}
\\
\hline
MIL-net   & \textbf{0.92} & 0.84&0.96&0.88\\
Baseline  & 0.87 & 0.83 & 1.0 & 0.90\\
Shallow-net & 0.50 & 0.58 & 1.0 & 0.73\\
ResNet-50  & 0.83 & 0.80 & 0.90 & 0.85\\
\hline
V-CNN  & 0.92 & 0.87 & 0.86 & 0.87 \\
SVM  & 0.77 & 0.87 & 0.86 & 0.77\\
\hline
\end{tabular}}

\end{table}

shallow CNN networks produced 75-90\% accuracy [16]. The performance of our proposed approach in detecting iPSC colony conditions is almost similar with those of the above studies. However the above mentioned stem cell detection studies used large series of training data. Whereas in this study we used limited number of dataset and different nature of stem cells than those of the above studies.

\subsection{Learning for localization}

\begin{figure*}
\centering
\includegraphics[width=0.85\textwidth]{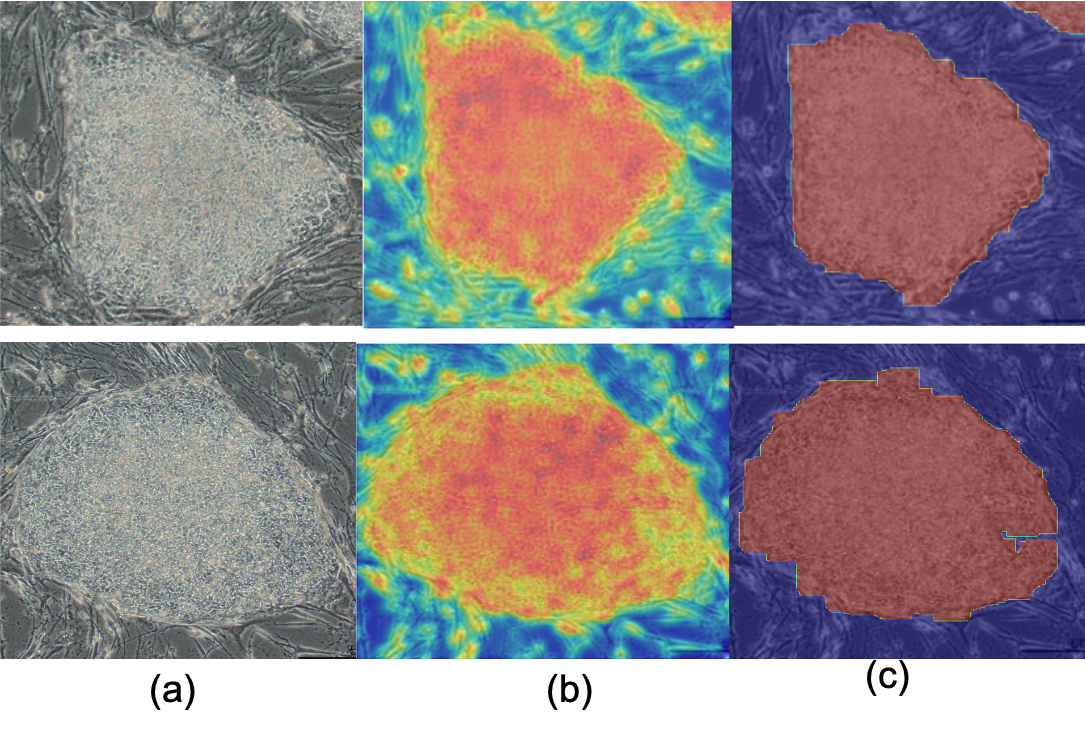}
\centering
\caption{\label{fig:ipscgood} Visualization of weak segmentation of good condition of colonies showing dense cells.(a) input image, (b) dense activation output, and (c) region of interest output}
\end{figure*}

\begin{figure*}
\centering
\includegraphics[width=0.85\textwidth]{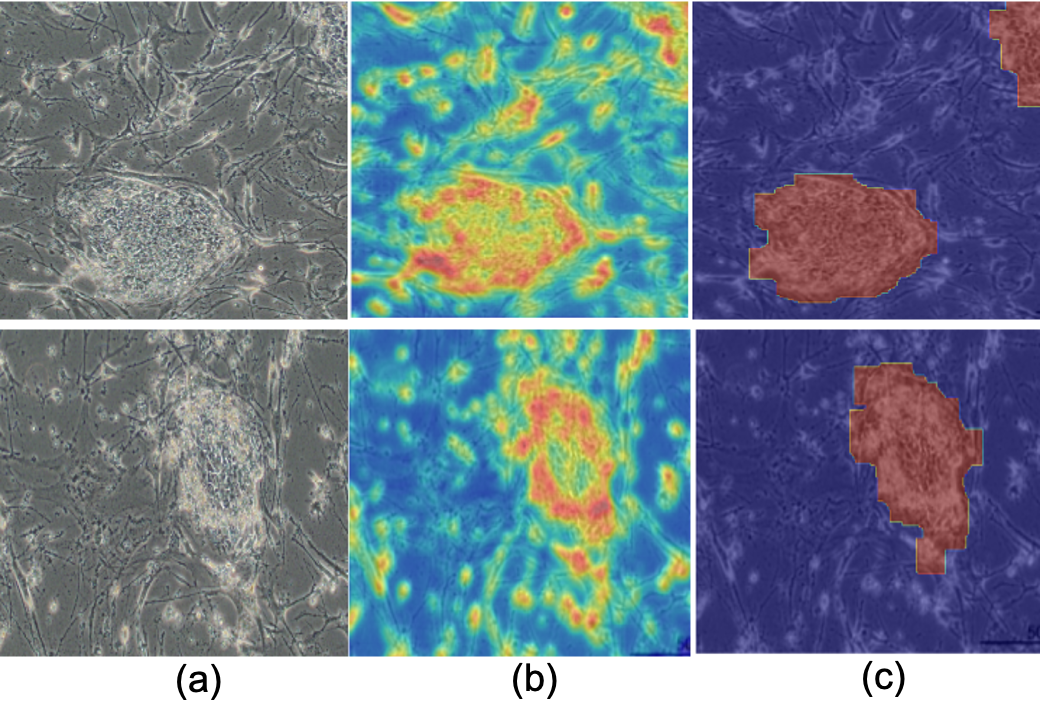}
\centering
\caption{\label{fig:ipscbad} Visualization of weak segmentation of bad condition of colonies showing sparse cells.(a) input image, (b) sparse activation output, and (c) region of interest output}
\end{figure*}

Though using limited number of training data, the decoder of the trained MIL-net can generate accurate activation of the region of interest. Figures \ref{fig:ipscgood} and \ref{fig:ipscbad} demonstrates the representative examples of the inferred  colony regions of the proposed network. The weekly localization of the cell colony can be clearly visualized from the texture features gathered from the average pooling of the MIL-net. The effect of MIL justified in this study is the extension of localization through the classification, though it is not directed to do so.

%$3 \times 3$ 

\section{Conclusions}

This study proposed a single network multiple instance learning in a weakly supervised settings based on U-net like architecture for annotating colonies and classifying the colony conditions. Most appealing in this study is the automatic visualization of the segmentation output of the cell regions without using the pixel-wise ground truth. Thus it reduced the annotation-cost and maximized the classification accuracy in detecting the colony conditions. Experimentally we proved the robustness of our proposed approach by comparing the performance with state-of-the-art methods. Furthermore, through the experiments, we observed that the proposed approach has fewer number of parameters and high detection ability when compared over CNN-based and SVM methods. Hence it indicated its simplicity and the reliability. Thus the proposed approach for extending the localization through classification is highly useful to explain the reasons for  decision making in identifying the colony conditions. Though our approach reveals high performance and produced inference-effective learning with weakly labels using iPSC dataset, the approach is needed to evaluate on different cell types data and different medical image dataset. It helps to understand the generalization ability of the proposed approach. In addition, the MIL-net with self-supervised  setting is needed to test the optimal procedure of the architecture.
 
\section*{conflict of interest}
The authors declare no conflicts of interest.

\section*{data availability statement}

Data available on request from the authors.

%\printendnotes

%\section{References}

% Submissions are not required to reflect the precise reference formatting of the journal (use of italics, bold etc.), however it is important that all key elements of each reference are included.
\bibliographystyle{IEEEtran}
\bibliography{sample}

\end{document}